\documentclass[aps,eqsecnum,nofootinbib]{revtex4}
\usepackage[usenames, dvipsnames]{color}
\usepackage{amssymb, amsmath, amsfonts, amsthm}
\usepackage[utf8]{inputenc} 
\usepackage{float}
\usepackage{flushend}
\usepackage{mathrsfs}
\usepackage{slashed}
\usepackage{bm}
\usepackage{chngcntr}
\usepackage{ytableau}
\usepackage{multirow}

\usepackage{chngcntr}

\counterwithin{table}{subsection} 

\newcommand{\be}{\begin{equation}}
\newcommand{\ee}{\end{equation}}
\newcommand{\bea}{\begin{eqnarray}}
\newcommand{\beas}{\begin{eqnarray*}}
\newcommand{\eea}{\end{eqnarray}}
\newcommand{\eeas}{\end{eqnarray*}}
\newcommand{\ba}{\begin{array}}
\newcommand{\ea}{\end{array}}

\newcommand{\ra}{\rangle}
\usepackage{hyperref}
\usepackage{url}
\hypersetup{
    colorlinks=true, 
    linkcolor=blue,
    filecolor=red,      
    urlcolor=cyan,
    citecolor=magenta}

\begin{document}

\title{Discrete Symmetries in the Chiral Fermion Formalism}

\author{I. P-Castro$^1$}
\email{ivan.perez.c@cinvestav.mx}

\author{J. L. Díaz-Cruz$^2$} 
\email{jldiaz@fcfm.buap.mx}

\author{A. Pérez-Lorenzana$^1$}
\email{aplorenz@fis.cinvestav.mx}

\affiliation{
$^1$Departamento de Física, Centro de Investigación y de 
Estudios Avanzados del I.P.N.\\ Apartado Postal 14-740, 07000, Mexico city, Mexico. \\
$^2$Facultad de Ciencias Físico-Matemáticas, Benemérita Universidad 
Autónoma de Puebla, Apartado Postal 1364, C.P. 72000, Puebla, Pue. México.
}

\begin{abstract}
In this paper, we present a revision of the discrete symmetries (C, P, T, CP, and CPT) within an approach that treats 2-component Weyl spinors as the fundamental building blocks. In particular, we show that we can define transformations for CP, T, and CPT without exchanging the right-handed and left-handed representative components that form a Dirac spinor, as is done in the traditional Dirac formalism. Then, we discuss some salient aspects of the discrete symmetries within quantum field theory (QFT). Besides the generic discussion, we also consider aspects arising within specific renormalizable theories, such as Quantum Electrodynamics (QED) and Yang-Mills (YM) theory. With the above, in addition to presenting a good introduction for non-experts, we establish how to use the chiral fermion formalism to study discrete symmetries in the context of QFT. 

\end{abstract}


\maketitle

\section{Introduction}
Discrete space-time symmetries—charge conjugation (C), parity (P), and time reversal (T)—along with combined symmetries, such as CP and CPT, play a crucial role in particle physics. The conservation of these symmetries imposes strict constraints on physical processes.

It has long been known that neither C, P, T, nor CP are always conserved at the quantum level. Fundamental interactions described by the Standard Model (SM) incorporate mechanisms that violate these symmetries, consistent with experimental observations. CPT, however, is believed to be conserved at all levels. This follows from the \textit{CPT theorem}, which states that any Lorentz-invariant, local quantum field theory with a Hermitian Lagrangian must preserve CPT symmetry (see a rigorous mathematical proof in~\cite{Streater2000pct}).

Most studies on discrete symmetries are formulated using standard four-component (Dirac-like) fermions. To account for the chiral nature of weak interactions, the explicit use of chiral projectors is often necessary. Although this approach is theoretically sound, it can appear cumbersome and less elegant, especially given the current development of two-component chiral fermion formalisms.

We present a comprehensive review of discrete symmetries in quantum field theory, including C, P, CP, T, and CPT, and offer an alternative approach that avoids ``mixing'' representative chiral components associated with irreducible representations of the Lorentz group in describing these transformations.

The organization of the paper is as follows:  
In Section \ref{Rev}, we provide a detailed discussion of discrete symmetries. We begin by revisiting C, P, and T within the canonical quantization framework, rederiving the corresponding transformation rules using the well-known four-component Dirac fermion formalism. Building on this, in the Section~\ref{D.symmetries} we introduce the chiral formalism and reexamine discrete space-time transformations in that language.

In Section~\ref{Chiral.transformations}, we focus on the properties of chiral currents under CPT transformations and establish their equivalence with the four-component formalism. Finally, in Section~\ref{More Examples}, we revisit CPT invariance in gauge field theories, with specific attention to QED and YM theories.
\section{Revisiting C, P, and T, with Dirac fermion formulation}\label{Rev}
We begin by providing a brief but general derivation for a 4-component fermion field transformation rules, using canonical quantization for guidance, for the three fundamental discrete space-time symmetries, C, P, and T. Interested readers can find further extended details on the literature (see for instance~\cite{Peskin2018introduction}). To this purpose, let us first introduce some general conventions. 
A 4-component fermion field, $\Psi$, is described by the Dirac equation
    \begin{equation}
        \left ( i\gamma^\mu\partial_\mu -m \right) \Psi(x) = 0 ~.
    \end{equation}
Here $\gamma_\mu$ stands for the four by four anticommuting Dirac matrices, which obey the algebra $\{\gamma^\mu,\gamma^\nu \} = 2 \eta^{\mu\nu}$, where the Minkowski metric is given by $(\eta^{\mu\nu}) = diag(1,-1,-1,-1)$. In the so-called chiral representation that we will use hereafter, gamma matrices are written as
    \begin{equation}
        \gamma^\mu = \left( \ba{cc} 
        0 & \sigma^\mu \\
        \bar\sigma^\mu & 0 
        \ea\right) ~,    
        \label{gamma}
    \end{equation}
where $\sigma^\mu= (\mathbf{1},\vec{\sigma})$ and $\bar\sigma^\mu= (\boldsymbol{1},-\vec{\sigma})$, and where $\sigma^0=\mathbf{1}$ is the two by two identity matrix and $\vec{\sigma}$ is the vector-matrix formed by the standard Pauli matrices,
    \begin{equation}
        \sigma^1 =\left(\ba{cc} 0&1\\1&0\ea\right)~;
        \sigma^2 =\left(\ba{cc} 0&-i\\i&0\ea\right)~;
        \sigma^3 =\left(\ba{cc} 1&0\\0&-1\ea\right)~.
    \end{equation}
Out of the four Dirac matrices, we define $\gamma_5 = i\gamma^0\gamma^1\gamma^2\gamma^3$ which anti-commutes with all other gamma matrices, $\{ \gamma^\mu,\gamma_5 \}=0$, and which in chiral representation becomes the diagonal matrix $\gamma_5=diag(-\mathbf{1},\mathbf{1})$.
This serves to separate the fermion component space into two subspaces, 
each one with eigenvalues $\mp 1$ under $\gamma_5$. These eigenvalues are referred to as chirality. The corresponding projection operators are the well-known left-handed,  $P_L = \frac{1}{2}(1-\gamma_5)$, and the right-handed, $P_R=\frac{1}{2}(1+\gamma_5)$, chiral projectors. In such terms, it is straightforward to write the fermion field as
    \begin{equation}
        \Psi(x) =\left(\ba{c} \varphi(x) \\ \bar\chi(x) \ea\right) ~,
        \label{Psi}
    \end{equation}
where the upper (lower) 2-fermion chiral component stands for the left (right) part of the Dirac field. Notice that we have introduced an overlined notation to distinguish the right-handed from left-handed components. It should not be confused with the usual bar operation on Dirac fermion fields, $\bar\Psi= \Psi^\dagger\gamma^0$. Due to the field mass, $m$, both the components are not independent, since the Dirac equation for them becomes,
    \begin{equation}
        i\sigma^\mu\partial_\mu \bar\chi= m \varphi \quad \text{and}\quad
        i\bar\sigma^\mu\partial_\mu \varphi= m \bar\chi \, .
    \end{equation}
From here, it is no difficult to see that for massless fermions, or equivalently at high energies, chiral states also become helicity eigenstates, associated with the helicity operator $h = \vec{\sigma}\cdot \vec{p}/|\vec{p}|$.

The standard solution to the Dirac equation in momentum space involves two spinorial wave functions, $u(\vec{p},s)$ and $v(\vec{p},s)$, associated with particle and antiparticle of momentum $p^\mu=(E,\vec{p})$ and spin projection $s$, respectively. The general mode expansion for a fermion operator with a given energy $E$ is then written, in canonical quantization, as
    \begin{equation}
        \Psi(x) = \sum_{s}\int \widetilde{d^3p} \big[ a(\vec{p}, s) u(\vec{p},s) e^{-i p\cdot x} + b^\dagger(\vec{p},s) v(\vec{p},s)e^{i p\cdot x} \big] ~.
        \label{psi}
    \end{equation}
where we have used the shorthand notation  $\widetilde{d^3p} = d^3p/2E(2\pi)^3$ and the sum is over all spin projection $s=\pm1$. Also, $a^\dagger(\vec{p},s)$ ($a(\vec{p},s)$) and $b^\dagger(\vec{p},s)$ ($b(\vec{p},s)$) stand for particle and antiparticle creation (annihilation) operators, which obey the anticommuting algebra given as 
$\left\{a(\vec{p},s),a(\vec{q},r)\right\} = \left\{b(\vec{p},s),b(\vec{q},r)\right\} = 0$, and $
\left\{a(\vec{p},s),b(\vec{q},r)\right\} =
\left\{a(\vec{p},s),b^\dagger(\vec{q},r)\right\} = 0$,
but $\left\{a(\vec{p},s),a^\dagger(\vec{q},r)\right\} = 
\left\{b(\vec{p},s),b^\dagger(\vec{q},r)\right\} = 
(2\pi)^32E\delta^{(3)}(\vec{p}-\vec{q}) \delta_{rs}$.
They act on particle number state space, where a single particle state of momentum $\vec{p}$ and spin projection $s$ is represented by $|\vec{p},s\ra$.
In what follows we will use these concepts to address the derivation for the transformation rules associated with C, P, and T.
\subsection{Charge conjugation}
Charge conjugation, C, is the quantum operation that replaces particles, by antiparticles without affecting space-time. For fermions, it is represented by the transformation of a given fermion operator into another one, which is described by the same Dirac equation but with the opposite charge. 
The change of the charge becomes evident once minimal coupling to gauge fields is included. 
This condition permits the definition of charge conjugation transformation as
    \begin{equation}
        {\cal C}^{-1} \Psi(x) {\cal C}  = D({\cal C})\bar\Psi^{*}(x) \, .
    \end{equation}
The action of ${\cal C}^{-1}$ on a given one particle state $|\vec{p},s\rangle$ should then reflect the transformation, mapping the state into that where the particle has the same momentum $\vec{p}$ and spin projection $-s$, $|\vec{p}, -s\rangle_{\rm c}$, but now the state corresponds to antiparticle. This implies for the creation operators that
    \begin{equation}
        {\cal C}^{-1} a^\dagger(\vec{p}, s) {\cal C} = n_c^{*} b^\dagger(\vec{p}, -s) \, , \quad \text{and} \quad {\cal C}^{-1} b^\dagger(\vec{p}, s) {\cal C} = n_c^\prime a^\dagger(\vec{p}, -s) \, ,
    \end{equation}
where $n_c^{(\prime)}$ are phase factors to be fixed. Therefore using the previous conditions in Eq.~\eqref{psi}, we gets
    \begin{equation*}
        {\cal C}^{-1}\Psi(x){\cal C} = 
        \sum_{s}\int \widetilde{d^3p} 
        \left[ n_c b(\vec{p}, -s) u^{*}(\vec{p},s) e^{i p\cdot x}
        + n_c^\prime a^\dagger(\vec{p},-s) v^{*}(\vec{p},s)e^{-i p\cdot x}\right] \, .
    \end{equation*}
Next, we use the following spinor wave function identities
    \begin{equation}
        i \gamma^2 u^{*}(\vec{p},s) = v(\vec{p},-s)\quad \text{and}\quad 
        -i\gamma^2 v^{*}(\vec{p},s) = u(\vec{p},-s) ~,
        \label{ctomp}
    \end{equation}
and chose the phase factors to be such that $n_c^\prime=n_c$, 
to rewrite the expression as
    \begin{equation}
        {\cal C}^{-1}\Psi(x){\cal C} = n_c i \gamma^2 \sum_{s}\int \widetilde{d^3p} 
        \left[ b(\vec{p}, s) v(\vec{p},s) e^{i p\cdot x}
        + a^\dagger(\vec{p},s) u(\vec{p},s)e^{-i p\cdot x}\right] \, ,
    \end{equation}
Because under the addition of the spin projector, we already have to take into account all the possibilities. Therefore, we identify $ D({\cal C}) = n_c i\gamma^2$. On the other hand, a Majorana fermion is then defined as that for which its corresponding fermion operator is an eigenstate of charge conjugation, which means that it should be chargeless and thus be its own antiparticle.
It is then simple to see that only one class of creation/annihilation operators is needed, which implies taking $b^\dagger(\vec{p},s) \rightarrow a^\dagger(\vec{p},s)$ within Eq.~\eqref{psi}, in their particular case.
Furthermore, considering the Majorana condition, the only possible physical choice for the phase factor is $n_c = \pm 1$. Here we take the convention $n_c = 1$, and thus C transformation on
fermions are expressed as
    \begin{equation}
        \Psi^c \equiv {\cal C}^{-1} \Psi {\cal C} = i\gamma^2 
        \Psi^* = C\bar\Psi^T = 
        i\gamma^2\gamma^0\bar\Psi^T ~.
        \label{DC}
    \end{equation}
Notice that with our conventions $C^{-1} = C^\dagger = C^T = -C$, and $C^\dagger \gamma^\mu C = -(\gamma^\mu)^T$. From this definition, it follows that $\bar\Psi^c\equiv{\cal C}^{-1} \bar \Psi {\cal C} = (\gamma^o 
\Psi^c)^\dagger= \Psi^T C$.
\subsection{P parity}
P parity is a discrete transformation associated to space reversion, where space-like coordinates, $\vec{x}$ are exchanged by $-\vec{x}$. 
As for any other transformation in space-time, this has consequences on the description of some physical quantities, such as exchanging the direction of the particle momentum but not that of angular momenta. Also, this would not modify the intrinsic properties of particles, like charge or spin. Of course, this last means that P should exchange helicity states since the quantity $\vec{p}\cdot\vec{s}$ would change its sign. As with charge conjugation, upon Dirac fermion field operators, the action of P parity can be represented by the overall action of a matrix, which is written as
    \begin{equation}
        {\cal P}^{-1} \Psi(x){\cal P} = D({\cal P})\Psi(Px) ~,
    \end{equation}
where the action on space-time, $Px$, is given by the matrix $({P^\mu}_\nu) = diag(1,-1,-1,-1)$. This implies that $D({\cal P})^2 = e^{i \phi}$, since the double transformation should map the fermion operator into itself, up to a general phase, $\phi$. 
The action of ${\cal P}^{-1}$ on a given one particle state $|\vec{p},s\rangle$ should then reflect the transformation, mapping the state into that where the particle has momentum $-\vec{p}$, but the same spin $s$, $|-\vec{p},s\rangle$. This implies for the creation operators that
    \begin{equation}
    \begin{aligned}
        {\cal P}^{-1} a^\dagger(\vec{p},s) {\cal P} = \epsilon^* a^\dagger(-\vec{p},s) \, , \quad \text{and} \quad
        {\cal P}^{-1} b^\dagger(\vec{p},s) {\cal P} &= \epsilon^\prime 
        b^\dagger(-\vec{p},s)~,
    \end{aligned}
    \end{equation}
where $\epsilon^{(\prime)}$ are phase factors to be fixed. Therefore, by 
using last condition jointly to Eq.~\eqref{psi} one gets that
    \begin{equation*}
        {\cal P}^{-1}\Psi(x){\cal P} = 
        \sum_{s}\int \widetilde{d^3p} 
        \left[\epsilon a(-\vec{p}, s) u(\vec{p},s) e^{-i p\cdot x}
        + \epsilon^\prime b^\dagger(-\vec{p},s) v(\vec{p},s)e^{i p\cdot x}\right]~.
    \end{equation*}
The last expression can be rewritten in a more convenient way by changing  
$\vec{p}\rightarrow-\vec{p}$ within the integral, to get
    \begin{equation*}
        {\cal P}^{-1}\Psi(x){\cal P} = \sum_{s}\int \widetilde{d^3p} \left[\epsilon a(\vec{p}, s) u(-\vec{p},s) e^{-i p\cdot Px} + \epsilon^\prime b^\dagger(\vec{p},s) v(-\vec{p},s)e^{i p\cdot Px}\right]~.
    \end{equation*}
Next, we use the following spinor wave function identities
    \begin{equation}
        \gamma^0 u(\vec{p},s) = u(-\vec{p},s)\quad \text{and}\quad 
        \gamma^0 v(\vec{p},s) = -v(-\vec{p},s) ~,
        \label{ptomp}
    \end{equation}
and chose the phase factors to be such that $\epsilon^\prime=-\epsilon$, to rewrite the expression as
    \begin{equation}
        {\cal P}^{-1}\Psi(x){\cal P} = \epsilon\gamma^0
        \sum_{s}\int \widetilde{d^3p} 
        \left[a(\vec{p}, s) u(\vec{p},s) e^{-i p\cdot Px}
        + b^\dagger(\vec{p},s) v(\vec{p},s)e^{i p\cdot Px}\right]
        = \epsilon\gamma^0\Psi(Px) ~.
    \end{equation}
Therefore, we identify $ D({\cal P}) = \epsilon \gamma^0$. Furthermore, considering that Majorana fermions should comply with the condition $b^\dagger(\vec{p},s) \rightarrow a^\dagger(\vec{p},s)$ then, the only possible physical choice for the phase factor is $\epsilon = \pm i$, since in such case above discussion requires that $\epsilon^*= -\epsilon$. Here we take the convention $\epsilon=i$, and thus P parity transformation on fermions is expressed as
    \begin{equation}
        {\cal P}^{-1}\Psi(x){\cal P} = D({\cal P})\Psi(Px)\equiv 
        i\gamma^0\Psi(Px) ~.
        \label{DP}
    \end{equation}
A straightforward calculation  also gives
    \begin{equation}
    \begin{aligned}
        {\cal P}^{-1}\bar\Psi(x){\cal P} &= 
        \left(\gamma^0D({\cal P})\Psi(Px)\right)^\dagger
        = \bar\Psi(Px)D({\cal P})^\dagger = -i\bar\Psi\gamma^0 ~. 
        \label{DPbar}
    \end{aligned}
    \end{equation}
Notice that $D({\cal P})=i\gamma^0$ is then a  unitary matrix, since $D({\cal P})^\dagger = D({\cal P})^{-1}$.
\subsection{Time reversal}
Wigner's Time reversal, ${\cal T}$ is given by an antilinear operation, which also acts on complex numbers as complex conjugation. In particular, it should be satisfied that ${\cal T}^{-1}~i~{\cal T} = -i$. Its action over a particle system should be that of reversing the evolution of time without affecting particle nature. In space-time, this is realized by the matrix operator $({T^\mu}_{\nu})=diag(-1,1,1,1)$.
Thus, on a particle state, the effect corresponds to that of inverting the direction of $\vec p$, as well as that of total angular momentum. Therefore, ${\cal T}$ should also invert the spin direction. Hence, for creation operators, we should now require that
    \begin{equation}
    \begin{aligned}
        {\cal T}^{-1} a^\dagger(\vec{p},s) {\cal T} = 
        \omega^*_s a^\dagger(-\vec{p},-s) \, , \quad \text{and} \quad
        {\cal T}^{-1} b^\dagger(\vec{p},s){\cal T} = \omega^\prime_s 
        b^\dagger(-\vec{p},-s) ~,
        \label{tat}
    \end{aligned}
    \end{equation}
where we have now introduced some general phase factors, $\omega^{(\prime)}_{s}$, which are  dependent on spin projection and remain to be fixed by requiring that the action on the general fermion operator be represented by a matrix mapping, 
    \begin{equation}
        {\cal T}^{-1}\Psi(x){\cal T} = D({\cal T}) \Psi(Tx)~.
    \end{equation}
The LHS of this last expression can be expanded using Eq.~\eqref{tat} and the action of ${\cal T}$ over complex numbers, as
    \begin{equation*}
        {\cal T}^{-1}\Psi(x){\cal T} = \sum_{s}\int \widetilde{d^3p} \left[\omega_s a(-\vec{p}, -s) u^*(\vec{p},s) e^{i p\cdot x} + \omega^\prime_s b^\dagger(-\vec{p},-s) v^*(\vec{p},s)e^{-i p\cdot x}\right]~.
    \end{equation*}
By redefining variables on this last expression, such that $\vec{p}\rightarrow 
-\vec{p}$ and $s\rightarrow -s$, we can write
$${\cal T}^{-1}\Psi(x){\cal T} = 
 \sum_{s}\int \widetilde{d^3p} 
\left[\omega_{-s} a(\vec{p}, s) u^*(-\vec{p},-s) e^{-i p\cdot Tx}
 + \omega^\prime_{-s} b^\dagger(\vec{p},s) v^*(-\vec{p},-s) 
 e^{i p\cdot Tx}\right]~.
$$
Next, we made use of the following identities,
    \bea
        \gamma_5 u(\vec{p},s) = s v(\vec{p},-s)~,~~~~~
        \gamma_5 v(\vec{p},s) = -s u(\vec{p},-s) \\
        v^*(\vec{p},s) = C\gamma^0 u(\vec{p},s)~,~~~~~
        u^*(\vec{p},s) = C\gamma^0 v(\vec{p},s)~,
    \eea
which jointly to those in Eq.~\eqref{ptomp} allow us to show that
    \begin{equation}
        u^*(-\vec{p},-s) = (-s)C\gamma_5u(\vec{p},s) \quad \text{and} \quad
        v^*(-\vec{p},-s) = (-s)C\gamma_5v(\vec{p},s)~,
    \end{equation}
and then to rewrite 
    $$
        {\cal T}^{-1}\Psi(x){\cal T} = 
        C\gamma_5 \sum_{s}\int \widetilde{d^3p} 
        \left[(-s)\omega_{-s} a(\vec{p}, s) u(\vec{p},s) e^{-i p\cdot Tx}
        + (-s)\omega^\prime_{-s} b^\dagger(\vec{p},s) v(\vec{p},s) 
        e^{i p\cdot Tx}\right]~.
    $$
Hence, we would require that $\omega_s = \omega_s^\prime = s\omega$, with $\omega$ a general phase factor. Nevertheless, in the case of Majorana fields, the same analysis will require that $\omega_s^* =\omega_s$. This would fix the overall phase factor to be $\omega= \pm 1$. Here we will take $\omega_s = \omega_s^\prime = s$, and thus the action of $\cal T$ over the general fermion operator shall be defined as
    \begin{equation}
        {\cal T}^{-1}\Psi(x){\cal T} = D({\cal T})\Psi(Tx)\equiv C\gamma_5\Psi(Tx) ~.
        \label{DT}
    \end{equation}
Notice that the matrix $D({\cal T})$ is also unitary. Similarly, we get 
    \begin{equation}
    \begin{aligned}
        {\cal T}^{-1}\bar\Psi(x){\cal T} &= 
        \left(\gamma^0C\gamma_5\Psi(Tx)\right)^\dagger = 
        \left(C\gamma_5\gamma^0\Psi(Tx)\right)^\dagger = \bar\Psi(Tx)\gamma_5C^\dagger=\bar\Psi(Tx) D({\cal T })^\dagger~.
    \end{aligned}
    \end{equation}
Notice also that $C\gamma_5 = i\gamma^2\gamma^0 (i\gamma^0\gamma^1\gamma^2\gamma^3)
=-\gamma^1\gamma^3$. Furthermore, for later calculations, one has to keep in mind that, due to the antilinear nature of T parity, matrix elements involving ${\cal T}$ are conjugated when acting on complex numbers.
\section{Discrete space-time symmetries with chiral fermion formalism} \label{D.symmetries}
Next, we review the chiral formalism and then rediscuss space-time discrete symmetries in this context. Let us first consider that, in the notation given by Eq.~\eqref{gamma}, fermion Lorentz algebra generators are given by
    \begin{equation}
        \frac{1}{2} \Sigma^{\mu\nu} = \frac{i}{4} [\gamma^\mu, \gamma^\nu] = \left( \ba{cc} i\sigma^{\mu\nu} & 0\\ 0 & i \bar\sigma^{\mu\nu} \ea \right) ~,
    \end{equation}
where $\sigma^{\mu\nu} = \frac{1}{4}(\sigma^\mu \bar\sigma^\nu - \sigma^\nu \bar\sigma^\mu)$ and $\bar\sigma^{\mu\nu} = \frac{1}{4}(\bar\sigma^\mu \sigma^\nu - \bar\sigma^\nu \sigma^\mu)$.  
Notice that the generator is diagonal by blocks. This means the 2-component fermions define two fundamental and non-equivalent, irreducible representations of the Lorentz group, i.e., proper Lorentz transformations do not mix them but map them within the same functional space. They are called chiral representations and are usually denominated as ${\cal F} =(\frac{1}{2},0)$ and $\bar{\cal{F}} = (0,\frac{1}{2})$, for the left and right handed fermion component spaces, respectively.

Under a  Lorentz transformation, $\Lambda$, they go as
    \begin{equation}
    \begin{aligned}
        \varphi(x) \rightarrow \varphi^\prime(x) = S(\Lambda)\, \varphi(\Lambda^{-1} x)
        \quad \text{but} \quad
        \bar\chi(x)\rightarrow \bar\chi^\prime(x) = 
        S(\Lambda)^{\dagger-1}\, \bar\chi(\Lambda^{-1} x)=
        S(\Lambda)^{-1 \dagger}\, \bar\chi(\Lambda^{-1} x)
        \label{lorentz}
    \end{aligned}
    \end{equation}
where we have used that $\bar\sigma^{\mu\nu} = -(\sigma^{\mu\nu})^\dagger $, and that $S(\Lambda)= \exp(\frac{1}{2}\omega_{\mu\nu}\sigma^{\mu\nu})$, with $\omega_{\mu\nu}$ the real transformation parameters.

The separation of the left-handed and right-handed fermion spaces makes it natural to ask about their properties under all other transformations associated to space-time. In order to address this, let us consider the most simple Lorentz invariant bilinear operator that can be formed from two 4-component fermions,
    \begin{equation}
        \Psi(x) =\left(\ba{c} \varphi(x) \\ \bar\chi(x) \ea\right)
        \qquad \text{and} \qquad
        \Phi(x) =\left(\ba{c} \lambda(x) \\ \bar\eta(x) \ea\right)~,
        \label{DFermions}
    \end{equation}
which, in terms of their chiral components, is given as 
$\bar\Psi\Phi = \bar\chi^\dagger\lambda + \varphi^\dagger\bar\eta$. Lorentz invariance means that each term in this expression must be a Lorentz invariant by itself, which is not difficult to see from Eq.~\eqref{lorentz}.

Hermitic conjugate operation, `$\dagger$', does not  map $\cal F$ into $\bar{\cal F}$. They are not the dual of each other. Dual spaces, denoted as $\cal F^*$ and $\bar{\cal F}^*$, respectively,  correspond to two new and independent irreducible representations of the Lorentz group. To enlighten the connection among $\cal F$ and $\bar{\cal F}$, we should notice that the 2-component fermion $\bar\varphi = i\sigma^2\varphi^*$ does actually transform as a fermion in $\bar{\cal F}$, since $\sigma^2\sigma^\mu\sigma^2 = \bar\sigma^{\mu T}$ and thus, $\sigma^2\sigma^{\mu\nu}\sigma^2 = - \sigma^{\mu\nu T}$. Conversely, $\chi = -i\sigma^2\bar\chi^*$ belongs to $\cal F$.
Furthermore, dual $\cal F^*$ ($\bar{\cal F}^*$) like fermions can also be obtained from $\cal F$ ($\bar{\cal F}$) like ones, from the mapping $\bar\chi^\dagger =  (i\sigma^2 \chi)^T$ [$\varphi^\dagger =  (-i\sigma^2 \bar\varphi)^T$], as their Lorentz transformation properties do show.
Hence, we can express the Lorentz invariants in terms of a very handy two-component notation where we take $\varphi_\alpha$ to denote $\cal F$ like fermions, and $\bar\chi^{\dot\alpha}$ for $\cal{\bar F}$ like ones, for $\alpha= 1,2$ and $\dot\alpha=\dot 1,\dot 2$, respectively. Therefore, we can write
    \begin{equation}
        \bar\chi^\dagger\lambda = (i\sigma^2\chi)^T\lambda =\chi^T(-i\sigma^2)\lambda
        =\epsilon^{\alpha\beta}\chi_\beta \lambda_\alpha ~,
    \end{equation}
and, similarly
    \begin{equation}
        \varphi^\dagger\bar\eta = (-i\sigma^2 \bar\varphi)^T\bar\eta = 
        \bar\varphi^T(i\sigma^2)\bar\eta = \epsilon_{\dot\alpha\dot\beta}\bar\varphi^{\dot\beta}\bar\eta^{\dot\alpha} ~,
    \end{equation}
where we have introduced the skew-symmetric matrices defined as, $\epsilon^{\alpha\beta} = (i\sigma^2)^{\alpha\beta}$ and $\epsilon_{\dot\alpha\dot\beta} = (-i\sigma^2)_{\dot\alpha\dot\beta}$. Notice that the index position is conventional to comply with our previous choice. 
It also becomes convenient to define the distinctive dual component notation as $\chi^\alpha = \epsilon^{\alpha\beta}\chi_\beta$ for $\cal F^*$ like fermions, and, similarly, $\bar\varphi_{\dot\alpha}= \epsilon_{\dot\alpha\dot\beta}\bar\varphi^{\dot\beta}$, for $\bar{\cal F}^*$ like ones. Thus, $\epsilon$ matrices can be conveniently used to lower and raise fermion indexes, which also amounts to mapping a fermion into dual. 
For this, we need also to define $\epsilon_{\alpha\beta} = (-i\sigma^2)_{\alpha\beta}$ and $\epsilon^{\dot\alpha\dot\beta} = (i\sigma^2)^{\dot\alpha\dot\beta}$, such that $\epsilon_{\alpha\beta}\epsilon^{\beta\rho}= {\delta_\alpha}^\rho$ and $\epsilon_{\dot\alpha\dot\beta}\epsilon^{\dot\beta\dot\rho}= 
{\delta_{\dot\alpha}}^{\dot\rho}$. 
With this notation, the invariants can now be rewritten in a simpler way, $\chi^\alpha \lambda_\alpha =\chi\lambda$ and  $\bar\varphi_{\dot\alpha}\bar\eta^{\dot\alpha} =\bar\varphi\bar\eta$.
It is then also straightforward to show that $\chi\lambda =\lambda\chi$ and that $\bar\varphi\bar\eta=\bar\eta\bar\varphi$ because the fermion fields anticommute. Finally, for the Dirac fermions, we then write
    \begin{equation}
        \bar\Psi\Phi = \left(\chi^\alpha~\bar\varphi_{\dot\alpha}\right)
        \left(\ba{c} \lambda_\alpha\\ \bar\eta^{\dot\alpha}\ea\right) = 
        \chi\lambda + \bar\varphi\bar\eta ~.
    \end{equation}
We can now proceed to our main topic of discussion. In what follows, in order to keep all things as clear as possible, we will be switching from Dirac to the chiral formulation as needed. Of course, we shall also use the Dirac formulation for discrete space-time transformations, as introduced in the previous section, to extract the corresponding rules that apply to chiral fermions.
\subsection{P parity}
By acting as indicated by Eq.~\eqref{DP}, we can easily see that P parity is a transformation that actually maps right-handed fermions into left-handed ones and vice versa, since
    \begin{equation}
        \Psi =\left(\ba{c} \varphi \\ \bar\chi \ea\right) 
        \,\xrightarrow{~~\text{P}~~}\,
        i\gamma_0 \Psi = \left(\ba{c} i\bar\chi\\ i\varphi \ea\right) ~,
        \label{Dpc}
    \end{equation}
where the action of the discrete transformations on space-time should be understood hereafter. Therefore, neither $\cal F$ nor $\bar{\cal F}$ 
are proper representations of P. Of course, that is also true for dual spaces. This will be a constant for all discrete transformations, as we shall show in any particular case.  
As a matter of fact only ${\cal F}\oplus \bar{\cal F}$ and 
${\cal F}^*\oplus\bar{\cal F}^*$ form complete irreducible representations for P. This will have an impact when trying to understand Dirac chiral currents involving $\gamma_5$, as we will discuss later. 

In terms of the chiral formalism above, transformation formally exchanges the nature of the index field without involving $\epsilon$ matrices. As we will see later, this is going to be a common feature for all discrete space-time transformations.
Therefore, as a way to keep formal track of the indexes we need to introduce a new conventional rule to account for this. Hereafter, 
we will use the identity matrix, $\sigma^0$, to express the index transformation by writing the mapping associated with a P as
    \begin{equation}
    \begin{aligned}
        \varphi_\alpha \,\xrightarrow{~\text{P}~}\,
        \bar\xi^{\dot\alpha}=
        i(\sigma^0)^{\dot\alpha\alpha}\varphi_\alpha~, \qquad \bar\varphi_{\dot\alpha} \,\xrightarrow{~\text{P}~}\, 
        \xi^\alpha =
        -i\bar\varphi_{\dot\alpha} (\sigma^0)^{\dot\alpha\alpha}~, \\
        \bar\chi^{\dot\alpha} \,\xrightarrow{~\text{P}~}\,
        \zeta_\alpha =   i(\sigma^0)_{\alpha\dot\alpha}\bar\chi^{\dot\alpha}~, \qquad \chi^{\alpha} \,\xrightarrow{~\text{P}~}\,
        \bar\zeta_{\dot\alpha}=
        -i\chi^{\alpha}(\sigma^0)_{\alpha\dot\alpha}~,
    \end{aligned} \label{p}
    \end{equation}
where we have introduced an explicit association to a given chiral spinor in the corresponding Lorentz representation as a reference to stress the position of the field in Dirac formalism and the nature of the expressions on the RHS of the equations. 

Although the above-given transformation rules are formally correct, they introduce some subtleties to properly reproduce the transformation properties of Dirac chiral currents. 
Consider for instance the chiral Lorentz invariant $\chi\lambda$. Under parity, according to the given rules, we will have that $\chi^\alpha\lambda_\alpha \, \xrightarrow{~\text{P}~} \, 
(-i\chi^\alpha(\sigma^0)_{\alpha\dot\alpha})(i(\sigma^0)^{\dot\alpha\beta}
\lambda_\beta) = \chi^\alpha\lambda_\alpha $.
Similarly, one can be shown that $\bar\varphi\bar\eta$ remains also invariant. 
Therefore, the scalar Dirac bilinear $\bar\Psi\Phi = \bar\varphi\bar\eta + \chi\lambda$ does remain also invariant under the action of P. That should also appear to be the case for the pseudo scalar current $\bar\Psi\gamma_5\Phi = \bar\varphi\bar\eta - \chi\lambda$.
However, in Dirac formalism, since $\{ \gamma^0,\gamma_5 \}=0$, we shall rather get, from Eq.~\eqref{Dcurrp}, that $\bar\Psi\gamma_5\Phi \, \xrightarrow{~\text{P}~}\, -\bar\Psi\gamma_5\Phi$. This apparent contradiction arises because, for four fermion components, parity is actually 
represented by the discrete operation $\varphi\leftrightarrow\bar\chi$, up to a global phase, which means that $\chi\lambda\leftrightarrow\bar\varphi\bar\eta$. 

To reconcile and clarify the issue, we go back to the field associations made in Eq.~\eqref{p}. There, it is clear that if one does make  explicit use of the associated field under P instead of the equivalence given in terms of the untransformed field components, then  chiral invariant current transformations should read as $\chi\lambda \, \xrightarrow{~\text{P}~} \,\bar\zeta_{(\chi)}\bar\xi_{(\lambda)}$ and 
$\bar\varphi\bar\eta \, \xrightarrow{~\text{P}~} \, \xi_{(\varphi)}\zeta_{(\eta)}$, with the field correspondence as indicated. 
In these terms, chiral Dirac current transformation would be written as $\bar\varphi\bar\eta - \chi\lambda \, \xrightarrow{~\text{P}~} \, -(\bar\zeta_{(\chi)}\bar\xi_{(\lambda)}-\xi_{(\varphi)}\zeta_{(\eta)})$, where now its RHS has the right expression form with the right sign. Further identification of the left and right-handed components to form four fermion fields that comply with what P does in Dirac formalism would remove the ambiguity. Since we only want to deal with fermions at the chiral level, our notation will always become ambiguous when dealing with P parity on chiral Dirac-like currents. In such cases, the use of Dirac formalism would always result in more transparency. That is a consequence of chiral formalism's explicit breaking of parity symmetry.
\subsection{Charge conjugation}
Once more, Lorentz's chiral representations are not complete representations under the action of charge conjugation. From Eq~\eqref{DC} we now see that
    \begin{equation}
        \Psi =\left(\ba{c} \varphi \\ \bar\chi \ea\right) \,\xrightarrow{~~\text{C}~~}\,
        \Psi^c=i\gamma^2\Psi^* = 
        -\left(\ba{c}  \chi \\ \bar\varphi \ea\right) ~. 
        \label{Dcc}
    \end{equation}
Therefore, C is also a transformation that maps $\cal F$ (${\cal F}^*$) into $\bar{\cal F}$ ($\bar{\cal F}^*$), and viceversa. In Dirac formalism, C has the particular effect of exchanging chiral components, plus the exchange of hand side on the resulting fermion. 
As before, from the last equation, we can read out the following formal expression for a C transformation on chiral formalism,
    \begin{equation}
        \ba{c} \varphi_\alpha \,\xrightarrow{~\text{C}~}\, -\bar\varphi^{\dot\alpha}~,
        \\[1em]
        \bar\chi^{\dot\alpha} \,\xrightarrow{~\text{C}~}\, -\chi_{\alpha}~,
        \ea
    \quad\text{and}\quad 
        \ba{c} \bar\varphi_{\dot\alpha} \,\xrightarrow{~\text{C}~}\, -\varphi^{\alpha}~,
        \\[1em]
        \chi^{\alpha} \,\xrightarrow{~\text{C}~}\,
        -\bar\chi_{\dot\alpha} ~.
        \ea
        \label{c}
    \end{equation}
As for parity, the above formal interpretation of the C action on the chiral fermions provides a source of ambiguity when dealing with chiral Dirac currents. The scalar contraction $\chi\lambda$ under C is mapped into $\bar\chi\bar\lambda = \bar\lambda\bar\chi$ whereas  $\bar\varphi\bar\eta$ goes to $\varphi\eta=\eta\varphi$. Therefore, for the Dirac scalar, we get that $\bar\Psi\Phi= \bar\varphi\bar\eta + \chi\lambda \, \xrightarrow{~\text{C}~} \, \eta\varphi+ \bar\lambda\bar\chi = \bar\Phi\Psi$ as expected. However, from the same rules, one gets that $\bar\Psi\gamma_5\Phi \, \xrightarrow{~\text{C}~} \, -\bar\Phi\Psi$, which has the opposite sign than expected. We then arise to a similar conclusion as for P. In regard to C transformation alone, the use of Dirac formalism results in being more appropriate to deal with theories that involve Dirac chiral currents, than chiral formalism.
\subsection{CP}
As already discussed above, both C and P, are unnatural in chiral formalism because chiral spaces do not form complete representations for such discrete transformation. We now explore the action of the combined action of CP. In terms of Dirac formalism, the action of CP is expressed as
    \begin{equation}
        \Psi =\left(\ba{c} \varphi \\ \bar\chi \ea\right) \,\xrightarrow{~\text{CP}~}\,\Psi^{CP}= -i C \Psi^* = - i\left(\ba{c} \bar\varphi \\ \chi \ea \right)~, 
        \label{Dcp}
    \end{equation}
as it can be easily checked from Eqs.~\eqref{Dpc} and \eqref{Dcc}. 
Notice that here there is no exchange of position among chiral fields. A left-handed field would remain to be left-handed. However, field content on the RHS of the above equation actually belong to a different representation, as their proper Lorentz transformations would show. 
Thus, CP maps $\cal F$ (${\cal F}^*$) like into $\bar{\cal F}$ ($\bar{\cal F}^*$) like fermions, and vice-versa.
Once more, to keep a careful track of the indexes, we can formally postulate CP action on chiral fermions as
    \begin{equation}
        \ba{c} \varphi_\alpha \,\xrightarrow{~\text{CP}~}\, 
        -i(\sigma^0)_{\alpha\dot\alpha}\bar\varphi^{\dot\alpha}~,
        \\[1em]
        \bar\chi^{\dot\alpha} \,\xrightarrow{~\text{CP}~}\, 
        -i(\sigma^0)^{\dot\alpha\alpha}\chi_{\alpha}~,
        \ea \, \text{and} \,
        \ba{c} \bar\varphi_{\dot\alpha} \,\xrightarrow{~\text{CP}~}\, 
        i\varphi^{\alpha}(\sigma^0)_{\alpha\dot\alpha}~,    
        \\[1em]
        \chi^{\alpha} \,\xrightarrow{~\text{CP}~}\,
        i\bar\chi_{\dot\alpha}(\sigma^0)^{\dot\alpha\alpha}~.
        \ea
        \label{cp}
    \end{equation}
Unlike C and P previously discussed transformations, CP ones become more natural for the chiral formalism. Consider the case of the Lorentz invariant $\chi\lambda$. Under CP, according to the above formal rules, it is transformed as follows,
    \begin{equation}
    \begin{aligned}
        \chi\lambda &= \chi^\alpha\lambda_\alpha \xrightarrow{~CP~} 
        i\bar\chi_{\dot\alpha}(\sigma^0)^{\dot\alpha\alpha}~
        (-i)(\sigma^0)_{\alpha\dot\beta}\bar\lambda^{\dot\beta} = 
        \bar\chi\bar\lambda~.
        \label{accp}
    \end{aligned}
    \end{equation}
Similarly, we get that $\bar\varphi\bar\eta \, \xrightarrow{~\text{CP}~} \, \varphi\eta$. Hence, for Dirac scalar currents we get that $\bar\Psi\Phi \, \xrightarrow{~\text{CP}~} \, \bar\Phi\Psi$, but $\bar\Psi\gamma_5\Phi \, \xrightarrow{~\text{CP}~} \, -\bar\Phi\gamma_5\Psi$, as expected. Unlike C and P alone, chiral formalism does not give rise to subtleties, becoming a natural framework for studying CP. We will explore other currents below.
\subsection{Time reversal}
From Eq.~\eqref{DT}, in terms of chiral components, we get 
    \begin{equation}
        \Psi =\left(\ba{c} \varphi \\ \bar\chi \ea\right) \,\xrightarrow{~~\text{T}~~}\,
        C\gamma_5\Psi = 
        \left(\ba{c} -i \sigma^2 \varphi\\ -i\sigma^2\bar\chi \ea\right) = \left( \ba{c} -\bar\varphi^{*} \\ \chi^{*}  \ea \right) ~.
        \label{Dtc}
    \end{equation}
As a quick inspection indicates,  the fermion components on the RHS of the above equation can be identified as chiral fields on dual representations.

Thus, we see that T transforms the chiral components into complex conjugated fields. In chiral formalism, complex conjugation is equivalent to dotting (undotting) the indexes, i.e. to map fields from $\cal F$ ( $\bar{\cal F}$ ) into $\bar{\cal F}^*$ ( ${\cal F}^*$ ), and vice-versa.
This is easy to see if one considers, for instance, that for $\varphi\in {\cal F}$, then $\bar\varphi = i\sigma^2\varphi^* \in \bar{\cal F}$. Hence, in terms of components we can write $(\varphi_\alpha)^{*} = \epsilon_{\dot\alpha\dot\beta}\bar\varphi^{\dot\beta} = 
\bar\varphi_{\dot\alpha}$.

Formal transformations in chiral formalism are then
    \begin{equation}
        \ba{c} \varphi_\alpha \,\xrightarrow{~\text{T}~}\, 
        -(\sigma^0)_{\alpha\beta}\varphi^{\beta}~,
        \\[1em]
        \bar\chi^{\dot\alpha} \,\xrightarrow{~\text{T}~}\, (\sigma^0)^{\dot\alpha\dot\beta}\bar\chi_{\dot\beta}~,
    \ea
        \quad\text{and}\quad 
        \ba{c} \bar\varphi_{\dot\alpha} \,\xrightarrow{~\text{T}~}\, 
        - \bar\varphi^{\dot\beta}(\sigma^0)_{\dot\beta\dot\alpha}~,
        \\[1em]
        \chi^{\alpha} \,\xrightarrow{~\text{T}~}\,
        \chi_{\beta}(\sigma^0)^{\beta\alpha} ~.
        \ea
        \label{t}
    \end{equation}
As before, let us consider the action of T on Lorentz scalars.
It is easy to see that under T we get $\chi^\alpha\lambda_\alpha \, \xrightarrow{~\text{T}~} \, -\chi_\alpha\lambda^\alpha =\lambda\chi$, and similarly we get $\bar\varphi\bar\eta \, \xrightarrow{~\text{T}~} \,  \bar\eta\bar\varphi$, which means that for Dirac four component fermions both $\bar\Psi\Phi$ and $\bar\Psi\gamma_5\Phi$ are indeed T invariants.
Like CP, chiral formalism is very well-suited for describing Dirac currents without ambiguities.
\subsection{CPT}
Finally, let us consider all discrete transformation combinations, referred to as CPT. From Dirac formalism definitions, given in Eqs.~\eqref{Dcp}, and~\eqref{Dtc} it is straightforward to see that
    \begin{equation}
        \Psi = \left(\ba{c} \varphi \\ \bar\chi \ea\right) \,\xrightarrow{~\text{T}~}\,  
        C \gamma_5 \Psi \xrightarrow{~\text{CP}~} i\gamma_5\Psi^* = \left(\ba{c} -i \varphi^*\\ i \bar\chi^* \ea\right)~,
        \label{Dtcp}
    \end{equation}
where we have made explicit use of $C^{*}=C=-C^{-1}$. Therefore, CPT transforms the chiral components into its complex conjugated fields up to a chiral phase. In chiral formalism, complex conjugation is equivalent to dotting (undotting) the indexes, i.e. to map fields from $\cal F$ ( $\bar{\cal F}$ ) into $\bar{\cal F}^*$ ( ${\cal F}^*$ ), and vice-versa. Of course, the identification can also be confirmed by its Lorentz transformation properties. By using this, we can write the formal chiral transformation rules for CPT as
    \begin{equation}
        \ba{c} \varphi_\alpha \,\xrightarrow{~\text{CPT}~}\, 
        -i{(\sigma^0)_{\alpha}}^{\dot\alpha}\bar\varphi_{\dot\alpha}~,
        \\[1em]
        \bar\chi^{\dot\alpha} \,\xrightarrow{~\text{CPT}~}\, 
        i{(\sigma^0)^{\dot\alpha}}_\alpha\chi^{\alpha}~,
        \ea \text{and}
        \ba{c} \bar\varphi_{\dot\alpha} \,\xrightarrow{~\text{CPT}~}\, 
        i\varphi_{\alpha}{(\sigma^0)^\alpha}_{\dot\alpha}~,
        \\[1em]
        \chi^{\alpha} \,\xrightarrow{~\text{CPT}~}\,
        -i\bar\chi^{\dot\alpha}{(\sigma^0)_{\dot\alpha}}^{\alpha}~.
        \ea
        \label{cpt}
    \end{equation}
In this section, we have introduced several conventions regarding phases and transformations in chiral formalism. However, we would like to emphasize that these conventions are not unique; other conventions can be established that lead to the same results discussed in later sections. For a more general discussion, see Chapter One of \cite{Dreiner2023spinors}, for example. The primary difference between our approach and that in the reference above is that we have avoided exchanging representative components of chiral fermionic fields in the transformation rules under discrete symmetries. This approach has introduced ambiguities in the transformations under P and C. However, no difficulties arise for CP, T, and CPT transformations.
\section{Chiral current transformations} \label{Chiral.transformations}
Using the decomposition of Dirac fields in terms of Weyl components given in Eq.~\eqref{DFermions}, we can rewrite the usual Dirac covariant bilinears in chiral formalism, as
    \begin{equation}
    \begin{aligned}
        \bar\Psi\Phi &= \bar\varphi\bar\eta + \chi\lambda \, ,\\
        \bar\Psi\gamma_5\Phi &= \bar\varphi\bar\eta - \chi\lambda \, ,\\
        \bar\Psi\gamma^\mu\Phi &= \chi\sigma^\mu\bar\eta 
        +\bar\varphi\bar\sigma^\mu\lambda \, , \\
        \bar\Psi\gamma^\mu\gamma_5\Phi &= \chi\sigma^\mu\bar\eta 
        -\bar\varphi\bar\sigma^\mu\lambda \, , \\
        \bar\Psi\Sigma^{\mu\nu}\Phi &= 2i\chi\sigma^{\mu\nu}\lambda 
        +2i \bar\varphi\bar\sigma^{\mu\nu}\bar\eta \, .
    \end{aligned}
    \end{equation}
A particular case of interest would be the free Dirac fermion Lagrangian, given by
    \begin{equation}
        {\cal L}_D=i\bar{\Psi}(x)\gamma^{
        \mu}\partial_{\mu}\Psi(x)-m\bar{\Psi}(x)\Psi(x)~.
        \label{Dcurr}
    \end{equation}
Notice that the derivative in the kinetic term does not affect the field transformations. Although it is itself transforms under P and T, but not under C, as $i\partial_\mu \, \xrightarrow{~\text{P}~} \, {P_\mu}^\nu i\partial_\nu$ and 
$i\partial_\mu \, \xrightarrow{~\text{T}~} \, -{T_\mu}^\nu i\partial_\nu$, respectively. 
Therefore, to build the standard Dirac fermion currents, one needs the following five fundamental covariant chiral bilinear currents,
    \begin{equation}
        \ba{rclcrcl}
        \chi\lambda &=& \chi^\alpha\lambda_\alpha = \lambda\chi &\hspace{2em}&
        \bar\varphi\bar\eta &=& \bar\varphi_{\dot\alpha}\bar\eta^{\dot\alpha} = 
        \bar\eta\bar\varphi \\
        \chi\sigma^\mu\bar\eta &=&
        \chi^\alpha(\sigma^\mu)_{\alpha\dot\alpha}\bar\eta^{\dot\alpha} = 
        -\bar\eta\bar\sigma^\mu\chi &&
        \bar\eta\bar\sigma^\mu\chi &=& 
        \bar\eta_{\dot\alpha}(\bar\sigma^\mu)^{\dot\alpha\alpha}\chi_{\alpha} \\
        i\chi\sigma^{\mu\nu}\lambda&=& 
        i\chi^\alpha(\sigma^{\mu\nu})_\alpha^{~\beta}\lambda_\beta = 
        -i\lambda\sigma^{\mu\nu}\chi &&
        i\bar\varphi\bar\sigma^{\mu\nu}\bar\eta &=& 
        i\bar\varphi_{\dot\alpha}(\bar\sigma^{\mu\nu})^{\dot\alpha}_{~\dot\beta}
        \bar\eta^ { \dot\beta } = 
        -i\bar\eta\bar\sigma^{\mu\nu}\bar\varphi~.
        \ea
    \end{equation}
Next, we shall explore explicitly their properties under the discrete space-time transformation, aiming to reproduce those of the Dirac formalism, as given in Eqs.~\eqref{Dcurrc},~\eqref{Dcurrp} and~\eqref{Dcurrt}.

As we have already stated in previous sections, from the analysis of the scalar and pseudo scalar Dirac currents in Eq.~\eqref{Dcurr},  C and P have an odd behavior in chiral formalism. Hence, we will not further extend the discussion of them here, but we will rather focus on CP, T, and CPT.

{\bf CP}. As we have already discussed previously, from Eq.~\eqref{accp}, under CP, the scalar chiral currents are exchanged into each other, $\chi\lambda\leftrightarrow\bar\chi\bar\lambda$, and so, it implies that $\bar\Phi\Psi \, \xrightarrow{~\text{CP}~} \, \bar\Phi\Psi$, whereas 
$\bar\Phi\gamma_5\Psi \, \xrightarrow{~\text{CP}~} \, \bar\Phi\gamma_5\Psi$. As for the vector current, we get
    \begin{equation}
    \begin{aligned}
        \chi\sigma^\mu\bar\eta = 
        \chi^\alpha(\sigma^\mu)_{\alpha\dot\alpha}\bar\eta^{\dot\alpha}
        \xrightarrow {~\text{CP}~} 
        &i\bar\chi_{\dot\beta}(\sigma^0)^{\dot\beta\alpha}
        (\sigma^\mu)_{\alpha\dot\alpha} (-i\sigma^0)^{\dot\alpha\beta}\eta_\beta = \bar\chi_{\dot\alpha}(\sigma^\mu)^{\dot\alpha\alpha}\eta_\alpha = -\eta_\alpha(\sigma^{\mu T})^{\alpha\dot\alpha}\bar\chi_{\dot\alpha}~.
    \end{aligned}
    \end{equation}
Using that $\sigma^{\mu T} = \sigma^2\bar\sigma^\mu\sigma^2$ and that $\bar\sigma^\mu = {P^\mu}_\nu\sigma^\nu$, above equation becomes
    \begin{equation}
    \begin{aligned}
        \chi\sigma^\mu\bar\eta 
        \xrightarrow {~\text{CP}~} 
        &-\eta_\alpha\epsilon^{\alpha\beta}
        (\bar\sigma^{\mu})_{\beta\dot\beta}\epsilon^{\dot\beta\dot\alpha}
        \bar\chi_{\dot\alpha} = -\eta^\beta(\bar\sigma^\mu)_{\beta\dot\beta}\bar\chi^{\dot\beta} = -{P^\mu}_\nu\eta\sigma^\nu\bar\chi~.
    \end{aligned}
    \end{equation}
It is then straightforward to see that $\bar\varphi\bar\sigma^\mu\lambda \, \xrightarrow{~\text{CP}~} \, -{P^\mu}_\nu~\bar\lambda\bar\sigma^\nu\varphi$, and therefore, for the four component fermions one gets that $\bar\Psi\gamma^\mu\Phi \, \xrightarrow{~\text{CP}~} \, -{P^\mu}_\nu~\bar\Phi\gamma^\mu\Psi$ and also that $\bar\Psi\gamma^\mu\gamma_5\Phi \, \xrightarrow{~\text{CP}~} \, -{P^\mu}_\nu~\bar\Phi\gamma^\mu\gamma_5\Psi$, as expected. A similar calculation shows for the tensor current that
    \begin{equation}
    \begin{aligned}
        i\chi\sigma^{\mu\nu}\lambda\xrightarrow{~\text{CP}~} 
        & i \bar\chi_{\dot\alpha} {(\sigma^{\mu\nu})^{\dot\alpha}}_{\dot\beta}
        \bar\lambda^{\dot\beta} = 
        -i \bar\lambda^{\dot\beta}{(\sigma^{\mu\nu T})_{\dot\beta}}^{\dot\alpha}
        \bar\chi_{\dot\alpha} = - i \bar\lambda_{\dot\beta} {(\sigma^{\mu\nu})^{\dot\beta}}_{\dot\alpha}
        \bar\chi^{\dot\alpha} =
        -i {P^\mu}_{\rho}{P^\nu}_{\tau}\bar\lambda\bar\sigma^{\rho\tau}\bar\chi ~,
    \end{aligned}
    \end{equation}
where the identities $\sigma^2\sigma^{\mu\nu}\sigma^2 = -\sigma^{\mu\nu T}$ and $\sigma^{\mu\nu}={P^\mu}_{\rho}{P^\nu}_{\tau}\bar\sigma^{\rho\tau}$ had been used. On the other hand, the algebra also shows that, 
$i\bar\varphi\bar\sigma^{\mu\nu}\bar\eta \, \xrightarrow{~\text{CP}~} \,
-i{P^\mu}_{\rho}{P^\nu}_{\tau}\eta\sigma^{\rho\tau}\varphi$.
As a direct implication, one gets $\bar\Psi\Sigma^{\mu\nu}\Phi \, \xrightarrow{~\text{CP}~} \,
-{P^\mu}_{\rho}{P^\nu}_{\tau}\bar\Phi\Sigma^{\rho\tau}\Psi$. 
Therefore, we get all CP transformation rules associated with fermion currents in accordance with Eqs.~\eqref{Dcurrc} and \eqref{Dcurrp}.

As an application, let us consider a free Dirac fermion, for instance, whose Lagrangian \eqref{Dcurr}, written in chiral formalism, is given as
    \begin{equation}
        {\cal L}_D = i\chi\sigma^\mu\partial_\mu\bar\chi 
        +i\bar\varphi\bar\sigma^\mu\partial_\mu\varphi  +
        m\bar\varphi\bar\chi + m\chi\varphi ~.
        \label{ld}
    \end{equation}
Under CP, the derivative is affected only by P, as $i\partial_\mu \, \xrightarrow{~\text{CP}~} \, {P_\mu}^\nu i\partial_\nu$, and thus the Lagrangian becomes
    \bea 
        {\cal L}_D &\xrightarrow{~\text{CP}~}&
        -{P^\mu}_\rho {P_\mu}^\nu \left[ i (\partial_\nu \chi)\sigma^\rho\bar\chi 
        + i(\partial^\nu\bar\varphi\bar) \sigma^\rho\varphi\right]  +
        m\varphi\chi + m\bar\chi\bar\varphi   \nonumber\\
        && =  i\chi\sigma^\nu\partial_\nu\bar\chi 
        +i\bar\varphi\bar\sigma^\nu\partial_\nu\varphi +
        m\bar\varphi\bar\chi + m\chi\varphi = {\cal L}_D ~,
    \eea
where ${P^\mu}_\rho {P_\mu}^\nu =\delta^\nu_\rho$ has been used, and a total derivative term has been neglected.

{\bf T}. Regarding time reversal, it has been shown above that the scalar currents remain invariant under T, and so do Dirac scalar and pseudo scalar bilinears $\bar\Phi\Psi$ and $\bar\Psi\gamma_5\Psi$. Vector currents, on the other hand, transform as follows,
    \begin{equation}
        \chi\sigma^\mu\bar\eta \,\xrightarrow{~\text{T}~}\, 
        \chi_{\beta}(\sigma^0)^{\beta\alpha}
        (\sigma^{\mu *})_{\alpha\dot\alpha} 
        (\sigma^0)^{\dot\alpha\dot\beta}\bar\eta_{\dot\beta} 
        = \chi_{\beta}(\sigma^{\mu *})^{\beta\dot\beta}\eta_{\dot\beta}~.
    \end{equation}
To write the RHS in a familiar way, we may use that $\sigma^{\mu *} = -{T^\mu}_\nu\bar\sigma^{\nu T}$ and thus one gets
    \begin{equation}
    \begin{aligned}
        \chi\sigma^\mu\bar\eta \,\xrightarrow{~\text{T}~}\, &-{T^\mu}_\nu\chi_{\beta} (\bar\sigma^{\nu T})^{\beta\dot\beta} 
        \bar\eta_{\dot\beta} = {T^\mu}_\nu\bar\eta_{\dot\beta}
        (\bar\sigma^\nu)^{\dot\beta\beta}\chi_{\beta} ={T^\mu}_\nu\bar\eta\bar\sigma^{\nu}\chi = -{T^\mu}_\nu\chi\sigma^{\nu}\bar\eta ~.
    \end{aligned}
    \end{equation}
This implies that $\bar\varphi\bar\sigma^\mu\lambda \, \xrightarrow{~\text{T}~} \, 
-{T^\mu}_\nu\bar\varphi\bar\sigma^\nu\lambda$ and for the Dirac currents that $\bar\Psi\gamma^\mu\Phi \, \xrightarrow{~\text{T}~} \, -{T^\mu}_\nu \bar\Psi\gamma^\nu\Psi$ and also that $\bar\Psi\gamma^\mu\gamma_5\Psi \, \xrightarrow{~\text{T}~} \,
-{T^\mu}_\nu\bar\Psi\gamma^\nu\gamma_5\Psi$. For tensor currents, we have
    \begin{equation}
    \begin{aligned}
        i\chi\sigma^{\mu\nu}\lambda\xrightarrow{~\text{T}~} 
        & i\chi_{\alpha} {(\sigma^{\mu\nu *})^{\alpha}}_{\beta}\lambda^{\beta} = 
        -i{T^\mu}_\rho{T^\nu}_\tau\chi_{\alpha}
        {(\sigma^{\rho\tau T})^{\alpha}}_{\beta}\lambda^{\beta} = 
        i{T^\mu}_\rho{T^\nu}_\tau\lambda^{\beta}
        {(\sigma^{\rho\tau})_{\beta}}^{\alpha}\chi_{\alpha}=
        -i{T^\mu}_\rho{T^\nu}_\tau\chi \sigma^{\rho\tau}\lambda ~,
    \end{aligned}
    \end{equation}
where the identity $\sigma^{\mu\nu *} = -{T^\mu}_\rho{T^\nu}_\tau\sigma^{\rho\tau T}$
has been used. Similarly, we also get $i\bar\varphi\bar\sigma^{\mu\nu}\bar\eta \, \xrightarrow{~\text{T}~} \, 
-i{T^\mu}_{\rho}{T^\nu}_{\tau}\bar\varphi\bar\sigma^{\rho\tau}\bar\eta$.
Therefore, for the Dirac current, we have that $\bar\Phi\Sigma^{\mu\nu}\Psi \, \xrightarrow{~\text{T}~} \, 
-{T^\mu}_{\rho}{T^\nu}_{\tau}\bar\Phi\Sigma^{\rho\tau}\Psi$. Hence, all Dirac current properties under T, as in Eq.~\eqref{DT}, are reproduced.

Next, let us consider once more the free fermion case in Eq.~\eqref{ld}. Now, under T, the mass terms $m\bar\varphi\bar\chi + m\chi\varphi$ remain explicitly invariant, leaving only to analyze the kinetic term, for which we have
    \begin{equation}
        i\chi\sigma^\mu\partial_\mu\bar\chi\xrightarrow{~\text{~\text{T}~}~}
        i{T^\mu}_\nu{T_\mu}^\rho\chi\sigma^{\nu}\partial_\rho\bar\chi = 
        i\chi\sigma^\nu\partial_\nu\bar\chi ~,
    \end{equation}
and also
    \begin{equation}
        i\bar\varphi\sigma^\mu\partial_\mu\varphi\xrightarrow{~\text{T}~}
        i{T^\mu}_\nu{T_\mu}^\rho \bar\varphi\sigma^{\nu}\partial_\rho\varphi = i\bar\varphi\sigma^\nu\partial_\nu\varphi~.
    \end{equation}
Here we have used that $i\partial_\mu \, \xrightarrow{~\text{T}~} \, -i{T_\mu}^\rho\partial_\rho$ and that ${T^\mu}_\nu{T_\mu}^\rho = \delta^\rho_\nu$. Hence, the kinetic terms are also invariant under T, and so does ${\cal L}_D$.

{\bf CPT}. Finally, let us address the transformation properties of the chiral currents under CPT. First, for the scalar currents, we can see, by using Eq.~\eqref{cpt}, that
    \begin{equation}
    \begin{aligned}
        \chi\lambda\xrightarrow{~\text{CPT}~} 
        & i\bar\chi^{\dot\alpha}{(\sigma^0)_{\dot\alpha}}^\alpha
        {(i\sigma^0)_\alpha}^{\dot\beta}\bar\lambda_{\dot\beta} =
        -\bar\chi^{\dot\alpha}\bar\lambda_{\dot\alpha} =
        \bar\lambda_{\dot\alpha}\bar\chi^{\dot\alpha} =
        \bar\lambda\bar\chi~,
        \end{aligned}
    \end{equation}
and also that $\bar\varphi\bar\eta \, \xrightarrow{~\text{CPT}~} \, \eta\varphi$. 
Furthermore, scalar and pseudo-scalar Dirac currents would transform as $\bar\varphi\bar\eta \pm \chi\lambda\rightarrow 
\eta\varphi \pm \bar\lambda\bar\chi =
\pm (\bar\lambda\bar\chi\pm\eta\varphi)$, just as expected from Eq.~\eqref{CPTtheo}.

For vector and tensor currents, we have to take into account that, inherited from T, under CPT one has that $({\cal CPT})^{-1} {\cal M} ({\cal CPT}) = {\cal M}^*$, for any matrix $\cal M$. In the case of vector ones, CPT field transformations indicate that
    \begin{equation}
    \begin{aligned}
        \chi\sigma^\mu\bar\eta \xrightarrow {~\text{CPT}~} 
        &i\bar\chi^{\dot\alpha}{(\sigma^0)_{\dot\alpha}}^{\alpha}
        (\sigma^{\mu *})_{\alpha\dot\beta} 
        {(-i\sigma^0)^{\dot\beta}}_{\beta}\eta^{\beta} 
        = \bar\chi^{\dot\alpha}(\sigma^{\mu T})_{\dot\alpha\beta}\eta^{\beta}
        = -\eta\sigma^\mu\bar\chi ~,
    \end{aligned}
    \end{equation}
and, similarly that $\bar\varphi\bar\sigma^\mu\lambda \, \xrightarrow{~\text{CPT}~} \, -\bar\lambda\bar\sigma^\mu\varphi$.
Tensor current, on the other hand, should transform as follows,
    \begin{equation}
    \begin{aligned}
        i\chi\sigma^{\mu\nu}\lambda\xrightarrow{~\text{CPT}~} 
        &i \chi^{\dot\alpha} {(\sigma^{\mu\nu *})_{\dot\alpha}}^{\dot\beta}
        \bar\lambda_{\dot\beta}= -i\bar\chi^{\dot\alpha}
        {(\bar\sigma^{\mu\nu T})_{\dot\alpha}}^{\dot\beta}\bar\lambda_{\beta} = 
        i\bar\lambda \bar\sigma^{\mu\nu}\bar\chi~,
    \end{aligned}
    \end{equation}
whereas $i\bar\varphi\bar\sigma^{\mu\nu}\bar\eta \, \xrightarrow{~\text{CPT}~} \, 
i\eta\sigma^{\mu\nu}\varphi$.
This implies that $\bar\Psi\gamma^\mu\Psi \, \xrightarrow{~\text{CPT}~} \, -\bar\Phi\gamma^\mu\Psi$,  
$\bar\Psi\gamma^\mu\gamma_5\Psi\, \xrightarrow{~\text{CPT}~} \, -\bar\Phi\gamma^\mu\gamma_5\Psi$
and $\bar\Psi\Sigma^{\mu\nu}\Psi \, \xrightarrow{~\text{CPT}~} \,\bar\Phi\Sigma^{\mu\nu}\Psi$,  
for the Dirac formalism, again, as expected.

Following above prescriptions, we see that Dirac mass term, under CPT, remains invariant, since now $\bar\varphi\bar\chi \leftrightarrow \chi\varphi$, whereas, for the kinetic term, using that $({\cal CPT})^{-1}i\partial_\mu ({\cal CPT}) = i\partial_\mu$, we get
    \begin{equation}
        i\chi\sigma^\mu\partial_\mu \bar\chi+i\bar\varphi\sigma^\mu\partial_\mu\varphi
        \xrightarrow{~\text{CPT}~}
        -i\partial_\mu\chi\sigma^\mu\bar\chi-i\partial_\mu\bar\varphi\sigma^\mu\varphi = 
        i\chi\sigma^\mu\partial_\mu         \bar\chi+i\bar\varphi\sigma^\mu\partial_\mu\varphi
        -i\partial_\mu (\chi\sigma^\mu\bar\chi +\bar\varphi\sigma^\mu\varphi) ~.
    \end{equation}
Therefore, up to a total derivative that does not contribute to the action, and which then can be neglected, this shows that we can consider the free Dirac fermion Lagrangian, ${\cal L}_D$ to be a CPT invariant.
\section{Discrete symmetries in Gauge Field Theories} \label{More Examples}
We will review some examples of gauge field theories in order to give an application of this formalism and clarify the appropriate transformations under discrete symmetries.
\subsection{QED}
The first and simplest example of a realistic theory with fermions is Quantum Electrodynamics, the theory of electrons and photons, which in chiral formalism is described by the Lagrangian
    \begin{equation}
    \begin{aligned}
        {\cal L}_{QED} &= -\frac{1}{4}F_{\mu\nu}F^{\mu\nu} 
        + i\ell_R\sigma^\mu D_\mu\bar \ell_R+
        i\bar\ell_L\bar\sigma^\mu D_\mu\ell_L - m\ell_R\ell_L -m\bar\ell_L\bar\ell_R~.
    \end{aligned}
        \label{qed}
    \end{equation}
This is also the simplest example of a Gauge Field Theory, where the covariant derivative that incorporates electron-to-photon interaction is defined through the minimal coupling rule, $D_\mu = \partial_\mu +i eA_\mu$, where $A_\mu$ stands for the electromagnetic four-vector potential, $e$ for the electron charge, $\ell_L$ ($\bar\ell_R$) for the left (right) handed electron field operator, and $F_{\mu\nu} = \partial_\mu A_\nu - \partial_\nu A_\mu$ for the electromagnetic field tensor. Let us briefly revise discrete transformations for this theory. The mass term has already been discussed in the previous section. To address all other terms, we need to specify the transformation rules for $A_\mu$.

Under charge conjugation, the exchange of particles by those with opposite charges affects electromagnetic potential, which is proportional to charge densities, by changing its global sign, that is ${\cal C}^{-1} A_\mu{\cal C} = -A_\mu$. As a consequence the covariant derivative changes as $D_\mu^c = {\cal C}^{-1} D_\mu{\cal C}  = \partial_\mu -i e A_\mu$, that can be read as the effective algebraic exchange of the charge sign.
The effect on the field tensor shall be the same as for the potential, ${\cal C}^{-1} F_{\mu\nu}{\cal C} = -F_{\mu\nu} $.

Regarding P parity, as this affects space-time, exchanging 
$\vec{x}\rightarrow -\vec{x}$, and thus changing the direction of velocities, $\vec{v}\rightarrow -\vec{v}$, and that of currents,  $\vec{j}\rightarrow -\vec{j}$, only the space like potential are affected in the same way. Hence, we can write ${\cal P}^{-1} A_\mu{\cal P} = {P_\mu}^\nu A_{\nu}$, where the transformed coordinate dependencies are to be understood, as usual. As we would also have ${\cal P}^{-1} \partial_\mu{\cal P} = {P_\mu}^\nu\partial_{\nu}$, for the covariant derivative we get that 
${\cal P}^{-1} D_\mu{\cal P}  = {P_\mu}^\nu D_\nu$, and so 
${\cal P}^{-1} F_{\mu\nu}{\cal P} = {P_\mu}^\rho{P_\mu}^\tau F_{\rho\tau}$. It is easy to see that the photon term, ${\cal L}_\gamma = -\frac{1}{4}F^2$, does remain invariant under any of those transformations. 
The invariance of the fermion theory under C and P, independently, has to be shown in the Dirac formalism, where they are well defined. Fermion mass and kinetic terms had also been previously shown to be invariant. Photon-to-fermion coupling, on the other hand, combines the transformation properties of the fermion vector current, as given in Eqs.~\eqref{DC} and \eqref{DP}, respectively, from where it is clear that
${\cal C}^{-1} \bar\Psi\gamma^\mu\Psi{\cal C} = -\bar\Psi\gamma^\mu\Psi$ and ${\cal P}^{-1} \bar\Psi\gamma^\mu\Psi{\cal P} = {P^\mu}_\rho 
\bar\Psi\gamma^\rho\Psi$ amount to keep action invariance. 

It is, on the other hand, illustrative to treat the combined action of CP in chiral formalism, for which we have $({\cal CP})^{-1} A_\mu({\cal CP}) = -{P_\mu}^\nu A_{\nu}$ and $({\cal CP})^{-1} D_\mu({\cal CP}) = -{P_\mu}^\nu D^c_\nu$, then $({\cal CP})^{-1} F_{\mu\nu}({\cal CP}) = -{P_\mu}^\rho{P_\mu}^\tau F_{\rho\tau}$.
The last expression indicates CP invariance of the photon kinetic term. 
The rest of the Lagrangian is transformed as follows, 
    \bea
        i\ell_R\sigma^\mu D_\mu\bar \ell_R+
        i\bar\ell_L\bar\sigma^\mu D_\mu\ell_L &\xrightarrow{~\text{CP}~}&
        -{P^\mu}_\nu{P_\mu}^\rho \left[i(D^c_\rho \ell_R)\sigma^\nu\bar\ell_R +
        i(D^c_\rho \bar\ell_L)\bar\sigma^\nu\ell_L \right]  \nonumber\\
        &&=
        -i (\partial_\nu\ell_R)\sigma^\nu\bar\ell_R - eA_\nu\ell_R\sigma^\nu\bar\ell_R
        -i (\partial_\nu\bar\ell_L)\bar\sigma^\nu\ell_L - 
        eA_\nu\bar\ell_L\bar\sigma^\nu\ell_L
        \nonumber \\
        &&=
        i\ell_R\sigma^\nu D_\nu \bar\ell_R + i\bar\ell_LD_\nu\bar\sigma^\nu\ell_L 
        -i\partial_\nu( \ell_R\sigma^\nu\bar\ell_R +\bar\ell_L\bar\sigma^\nu\ell_L) ~,
    \eea
which shows that the whole action remains CP invariant.

Time reversal change, the global sign of velocities and momenta, and thus also that of space like current sources, so that ${\cal T}^{-1}j_\mu{\cal T} = -{T_\mu}^\nu j_\nu$. This is physically reflected in the electromagnetic potential, ${\cal T}^{-1}A_\mu{\cal T} = -{T_\mu}^\nu A_\nu$. As time reversal also conjugates all complex numbers, we get for the covariant derivative that ${\cal T}^{-1}iD_\mu{\cal T} = -{T_\mu}^\nu iD_\nu$. Hence, the electromagnetic tensor transforms as
    \begin{equation}
        {\cal T}^{-1}F_{\mu\nu}{\cal T} = -{T_\mu}^\rho {T_\nu}^\tau F_{\rho\tau} \, ,
    \end{equation}
which leaves ${\cal L}_\gamma$ invariant. For fermion terms, a simple calculation also shows that
    \begin{equation}
        i\ell_R\sigma^\mu D_\mu\bar \ell_R + i\bar\ell_L\bar\sigma^\mu D_\mu\ell_L  
        \xrightarrow{~\text{T}~}
        -{T^\mu}_\nu \left[\ell_R\sigma^\nu (-i{T_\mu}^\rho D_\rho)\bar\ell_R 
        + \bar\ell_R\bar\sigma^\nu (-i{T_\mu}^\rho D_\rho)\ell_L \right]
        = 
        i\ell_R\sigma^\nu D_\nu\bar\ell_R + i\bar\ell_L\bar\sigma^\nu D_\nu\ell_L ~.
    \end{equation}
Finally, by combining all above-given rules, we can see that the vector potential transforms under CPT simply as $({\cal CPT})^{-1}A_\mu({\cal CPT}) = - A_\mu$, whereas $({\cal CPT})^{-1}i\partial_\mu({\cal CPT}) =  i\partial_\mu$, which means that $({\cal CPT})^{-1}iD_\mu({\cal CPT}) =  iD^c_\mu$ but $({\cal CPT})^{-1}F_{\mu\nu}({\cal CPT}) = F_{\mu\nu}$. Hence, ${\cal L}_\gamma$ is explicitly invariant, whereas the fermion terms remain invariant up to a total derivative,
        \bea
        i\ell_R\sigma^\mu D_\mu\bar \ell_R+
        i\bar\ell_L\bar\sigma^\mu D_\mu\ell_L &\xrightarrow{~\text{CPT}~}&
        -i(D^c_\mu\ell_R)\sigma^\mu\bar\ell_R +
        i(D^c_\mu\bar\ell_L)\bar\sigma^\mu\ell_L \nonumber\\
        &&= 
        i\ell_R\sigma^\mu D_\mu \bar\ell_R + i\bar\ell_LD_\mu\bar\sigma^\nu\ell_L 
        -i\partial_\mu( \ell_R\sigma^\mu\bar\ell_R +\bar\ell_L\bar\sigma^\mu\ell_L) ~.
    \eea
That warrants CPT invariance of QED action. 
\subsection{Yang-Mills theories}
In Gauge Field Theories, left- and right-handed fermions can belong to different representations of the gauge group. As a consequence C, and P are no anymore conserved.  Such is the case of the electroweak theory, for instance. 
For them, chiral formulation becomes natural. In order to study other discrete symmetries in this context one has to generalize our previous results. To be specific, without lack of generality, let us consider a gauge field theory based on the compact gauge group, $G$, where left (right) handed fermions are assigned to an irreducible representation generically denoted as $\varphi$ ($\bar\chi$), of dimension $n_L$ ($n_R$). The covariant derivative shall now involve the corresponding  (hermitian matrix) representations of the  generators of $G$, denoted as $T^a_{L,R}$ respectively, such that
    \begin{equation}
        i\boldsymbol{D}_{L\mu}\varphi = 
        (i\partial_\mu - g{\boldsymbol{A}}_{L\mu})\varphi~, \quad \text{but} \quad
        i\boldsymbol{D}_{R\mu} \bar\chi = 
        (i\partial_\mu - g{\boldsymbol{A}}_{R\mu})\bar\chi ~,
    \end{equation}
where ${\boldsymbol{A}}_{L,R\mu} = A_{a\mu}T^a_{L,R}$, and $A_{a\mu}$ stands for the gauge fields, and $g$ for the gauge coupling constant. In any representation, group generators are normalized by the condition $Tr(T^bT^b) = \frac{1}{2}\delta^{ab}$ and obey the Lie group algebra $[T^a,T^b] = if^{abc}T^c$, where $f^{abc}$ are the skewsymmetric structure constants.
Gauge fields form by themselves an irreducible representation of $G$, called the adjoint representation.
The  covariant gauge field tensor is then defined using adjoint representation as
    \begin{equation}
        \boldsymbol{F}_{\mu\nu} = \frac{i}{g}[i\boldsymbol{D}_\mu,i\boldsymbol{D}_\nu] 
        = \partial_\mu\boldsymbol{A}_\nu-\partial_\nu\boldsymbol{A}_\mu
        +ig[\boldsymbol{A}_\mu,\boldsymbol{A}_\nu] ~.
    \end{equation}
Non-abelian generalization of QED Lagrangian, in Eq.~\eqref{qed}, then reads as
    \begin{equation}
        {\cal L}_{GFT} = -\frac{1}{2}Tr\boldsymbol{F}_{\mu\nu}\boldsymbol{F}^{\mu\nu} 
        + i\chi\sigma^\mu \boldsymbol{D}_{R\mu}\bar \chi+
        i\bar\varphi\bar\sigma^\mu \boldsymbol{D}_{L\mu}\varphi~.
        \label{gft}
    \end{equation}
Mass terms are, in general, non-gauge invariant. In the SM, they are generated in a gauge covariant way by the Higgs mechanism. In our generic model, that would require including a proper set of complex scalar fields, $\phi$, in any of the irreducible representations contained in the $n_R\times n_L$ representation. This would allow us to write the following additional terms, including Yukawa couplings,
    \begin{equation}
        {\cal L}_{\phi} + {\cal L}_{Y} = 
        (\boldsymbol{D}^\mu\phi)^\dagger \boldsymbol{D}_\mu\phi - V(\phi^\dagger\phi) - 
        \left(y \chi\phi\varphi + y^\dagger\bar\varphi\phi^\dagger\bar\chi\right)~,
    \end{equation}
where generation and group representation indexes are to be understood. Also, $V$ stands for the gauge invariant scalar potential.
For one-generation flavorless models, the Yukawa couplings matrix, $y$, reduces to a constant that can always be taken to be real. In general, as in the SM, they are complex matrices that can induce CP violation due to flavor mixings. 

In order to discuss CP, we need to first consider C and P transformation rules for gauge and scalar fields. Scalars, by definition, are P invariants, ${\cal P}^{-1}\phi{\cal P}=\phi$, whereas gauge fields have a proper vector-like transformation, ${\cal P}^{-1}{\boldsymbol{A}}_\mu{\cal P}={P_{\mu}}^\nu{\boldsymbol{A}}_\nu$, and thus, as for the abelian QED case, we also have ${\cal P}^{-1}{\boldsymbol{D}}_\mu{\cal P}=
{P_{\mu}}^\nu{\boldsymbol{D}}_\nu$ and ${\cal P}^{-1}{\boldsymbol{F}}_{\mu\nu}{\cal P} ={P_{\mu}}^\rho{P_{\nu}}^\tau{\boldsymbol{F}}_{\rho\tau}$.

Charge conjugation, on the other hand, acts on scalar representations as complex conjugation,  ${\cal C}^{-1}\phi{\cal C}=\phi^*$. On gauge fields, its action is given as
    \begin{equation}
        {\cal C}^{-1} {\boldsymbol{A}}_\mu{\cal C} = -{\boldsymbol{A}}_\mu^T~.
    \end{equation}
This means that gauge fields associated with symmetric (antisymmetric) generators are odd (even) under C. Therefore, for the covariant derivate, we get ${\boldsymbol{D}}_\mu^c = {\cal C}^{-1}{\boldsymbol{D}}_\mu{\cal C} = \partial_\mu - ig{\boldsymbol{A}}_\mu^T$ and for the gauge field tensor $ {\cal C}^{-1}{\boldsymbol{F}}_{\mu\nu}{\cal C} = -{\boldsymbol{F}}_{\mu\nu}^T$. 
Hence, under the combined action of CP, one has that
$\phi\rightarrow\phi^*$, ${\boldsymbol{A}}_\mu\rightarrow -{P_{\mu}}^\nu{\boldsymbol{A}}_\nu^T$, ${\boldsymbol{D}}_\mu\rightarrow {P_{\mu}}^\nu{\boldsymbol{D}}_\nu^c$ and also ${\boldsymbol{F}}_{\mu\nu}\rightarrow -{P_{\mu}}^\rho{P_{\nu}}^\tau{\boldsymbol{F}}_{\rho\tau}^T$.
It is then easy to see that ${\cal L}_{YM}=-\frac{1}{2} Tr{\boldsymbol{F}}^2$ and ${\cal L}_\phi$ are C and P, and so CP, invariants. In particular, one sees that
    \begin{equation}
    \begin{aligned}
        (\boldsymbol{D}^\mu\phi)^\dagger \boldsymbol{D}_\mu\phi\xrightarrow{~\text{C}~}
        & (\boldsymbol{D}^{c\mu}\phi^*)^\dagger \boldsymbol{D}^c_\mu\phi^* = (\boldsymbol{D}^\mu\phi)^T (\boldsymbol{D}_\mu^\dagger)^T(\phi^\dagger)^T = (\boldsymbol{D}^\mu\phi)^\dagger \boldsymbol{D}_\mu\phi ~.
    \end{aligned}
    \end{equation}
Yukawa couplings transform under CP as $y\chi\phi\varphi\rightarrow y\bar\chi\phi^*\bar\varphi = 
y^T\bar\varphi\phi^\dagger\bar\chi$, which shows that CP is violated unless the couplings are real numbers.
Fermion lagrangian terms are also invariant, up to a total derivative, since
    \begin{equation}
        i\chi\sigma^\mu{\boldsymbol{D}}_{R\mu}\bar\chi \xrightarrow{~\text{CP}~}
        i{P_\mu}^\nu{P^\mu}_\rho\bar\chi_{\dot\alpha}
        (\bar\sigma^\rho)^{\dot\alpha\alpha}
        \left( \partial_\nu -ig{\boldsymbol{A}}_{R\nu}^T\right)\chi_\alpha = 
        i\chi \sigma^\nu \left( \partial_\nu 
        +ig{\boldsymbol{A}}_{R\nu}\right)\bar\chi_{\dot\alpha}  
        -i\partial_\nu(\chi\sigma^\nu\bar\chi)~,
    \end{equation}
and
    \begin{equation}
        i\bar\varphi\bar\sigma^\mu{\boldsymbol{D}}_{L\mu}\varphi \,\xrightarrow{~\text{CP}~}\, i{P_\mu}^\nu{P^\mu}_\rho\varphi^{\alpha}
        (\sigma^\rho)_{\alpha\dot\alpha}
        \left( \partial_\nu -i 
        g{\boldsymbol{A}}_{L\nu}^T\right)\bar\varphi^{\dot\alpha} = 
        i\bar\varphi\bar\sigma^\nu 
        \left( \partial_\nu+ig{\boldsymbol{A}}_{L\nu}\right)\varphi  
        -i\partial_\nu(\bar\varphi\bar\sigma^\nu\varphi)~.
    \end{equation}
Next, let us discuss time reversal. First, scalar fields are invariant under T. Therefore, the action of T on Yukawa couplings goes as $y\chi\phi\varphi\rightarrow y^*\chi\phi\varphi$, which again shows that, in general, they violate T symmetry, provided the couplings are non-real Gauge fields, on the other hand, transform according to the rule
    \begin{equation}
        {\cal T}^{-1}{\boldsymbol{A}}_\mu {\cal T} = -{T_\mu}^\nu{\boldsymbol{A}}_\nu~,
    \end{equation}
which means that $A_{a\mu}$ is odd (even) under T if the associated 
group generator is real (pure imaginary). From here, it comes out that 
${\cal T}^{-1}i{\boldsymbol{D}}_\mu{\cal T} = -{T_\mu}^\nu 
i{\boldsymbol{D}}_\nu$ and thus, that 
${\cal T}^{-1}{\boldsymbol{F}}_{\mu\nu}{\cal T} = 
-{T_\mu}^\rho {T_\mu}^\tau  {\boldsymbol{F}}_{\rho\tau}$.
This implies that ${\cal L}_{YM}$ is a T invariant, as well as ${\cal L}_\phi$.
Fermion terms now transform as
    \begin{equation}
    \begin{aligned}
        i\chi\sigma^\mu{\boldsymbol{D}}_{R\mu}\bar\chi \,\xrightarrow{~\text{T}~}\,
        &-i{T_\mu}^\nu\chi_{\alpha}
        (\sigma^{\mu*})^{\alpha\dot\alpha}{\boldsymbol{D}}_{R\nu}\bar\chi_{\dot\alpha} = i{T_\mu}^\nu {T^\mu}_\rho \chi^\alpha (\sigma^\rho)_{\alpha\dot\alpha}  {\boldsymbol{D}}_{R\nu}\bar\chi^{\dot\alpha} =
        i\chi \sigma^\nu{\boldsymbol{D}}_{R\nu}\bar\chi ~,
    \end{aligned}
    \end{equation}
where the identity $\sigma^{\mu*} = -{T^\mu}_\rho\sigma^2\sigma^{\rho}\sigma^2$ 
has been used. A similar algebra would also show the invariance of the left-handed term, $i\bar\varphi\bar\sigma^\mu \boldsymbol{D}_{L\mu}\varphi$.

By using the above results, one obtains that, under CPT, 
$\phi\rightarrow\phi^*$, 
${\boldsymbol{A}}_\mu\rightarrow -{\boldsymbol{A}}_\mu^T$, 
$i{\boldsymbol{D}}_\mu\rightarrow i{\boldsymbol{D}}_\nu^c$ and 
$ {\boldsymbol{F}}_{\mu\nu}\rightarrow {\boldsymbol{F}}_{\mu\nu}^T$. 
The invariance of the whole Lagrangian under these transformations will easily follow from a simple algebra. In particular, it is interesting to notice that for Yukawa couplings the combined CPT action,
$y\chi\phi\varphi\xrightarrow{~\text{T}~} y^*\chi\phi\varphi
\xrightarrow{~\text{CP}~} y^\dagger\bar\varphi\phi^\dagger\bar\chi$,
maps each term into its hermitic conjugated one. Thus, CPT invariance follows from the real nature of the Lagrangian, regardless of the fact that these same terms break CP and T independently.

\section{Conclusions and Outlook}
In this paper, we have reviewed the discrete symmetries: C, P, T, CP, and CPT. We discuss the fundamental aspects of these symmetries using the standard formalism of four-component Dirac spinors, thereby establishing our conventions and definitions for the chiral fermion formalism.

We point out the lack of clarity that often arises when studying symmetries like C and P in the chiral formalism without exchanging representative chiral fields. However, we show that treating CP, T, and CPT symmetries in the same formalism is more natural and does not require exchanging representative chiral fields.

This approach provides a systematic way to study discrete symmetries in any QFT. As an example, we analyze relevant aspects in specific renormalizable theories, such as quantum QED and YM theories. We believe this methodology can be particularly useful for students and non-specialist researchers. Moreover, as we show in subsequent research, this formalism is well-suited for studying CPT violation in extensions of the Standard Model using the effective field theory approach.

\acknowledgments
JLDC acknowledges support from SNII (CONAHCYT) and VIEP (BUAP). APL wants to thank FCFM-BUAP for the warm and enjoyable hospitality during his sabbatical leave and where part of this work was done. The work of APL has been partially supported by CONAHCYT, Mexico, under grant 237004. IPC acknowledges financial support from CONAHCYT, Mexico, as ``Ayudante SNII III''.
\appendix
\section{General current bilinear transformation rules}
Using previously given transformation rules for general four-component fermions, it is straightforward to calculate the corresponding transformation rules of main bilinear, Lorentz covariant, currents. They are as follows.

{\bf Under ${\cal C}$.} Charge conjugation, acting on a general bilinear current, is expressed as ${\cal C}^{-1}\bar\Psi {\cal M}\Phi{\cal C} = 
{\cal C}^{-1}\bar\Psi{\cal C} {\cal M}{\cal C}^{-1}\Phi{\cal C} = \Psi^T C{\cal M}C\bar\Phi^T = -\bar\Phi C^T{\cal M}^TC^T\Psi = \bar\Phi C^\dagger{\cal M}^TC\Psi$. Therefore, we get
    \bea
        {\cal C}^{-1}\bar\Psi\Phi{\cal C} &=&  
        +\bar\Phi\Psi~,\nonumber\\
        {\cal C}^{-1}\bar\Psi \gamma_5\Phi{\cal C} &=&  
        +\bar\Phi \gamma_5\Psi~,\nonumber\\
        {\cal C}^{-1}\bar\Psi\gamma^\mu\Phi{\cal C} &=&  
        -\bar\Phi\gamma^\mu\Psi~, \nonumber\\
        {\cal C}^{-1}\bar\Psi\gamma^\mu\gamma_5\Phi{\cal C} &=&  
        +\bar\Phi\gamma^\mu\gamma_5\Psi~, \nonumber\\
        {\cal C}^{-1}\bar\Psi\Sigma^{\mu\nu}\Phi{\cal C} &=&  
        -\bar\Phi\Sigma^{\mu\nu}\Psi~.
        \label{Dcurrc}
    \eea
{\bf Under $\cal P$}. For the scalar current, we have that 
$${\cal P}^{-1}\bar\Psi(x)\Phi(x){\cal P} =  
{\cal P}^{-1}\bar\Psi(x){\cal P}{\cal P}^{-1}\Phi(x){\cal P} = \bar\Psi(Px)D({\cal P})^\dagger D({\cal P})\Phi(Px) =  
+\bar\Psi(Px)\Phi(Px)~,$$ 
and similarly, for the other four fundamental covariant currents, we get
    \bea
        {\cal P}^{-1}\bar\Psi(x)\gamma_5\Phi(x){\cal P} &=&  
        -\bar\Psi(Px)\gamma_5\Phi(Px)~,\nonumber\\
        {\cal P}^{-1}\bar\Psi(x)\gamma^\mu\Phi(x){\cal P} &=&  
        +P^\mu_{~\nu}\bar\Psi(Px)\gamma^\nu\Phi(Px)~, \nonumber\\
        {\cal P}^{-1}\bar\Psi(x)\gamma^\mu\gamma_5\Phi(x){\cal P} &=&  
        -P^\mu_{~\nu}\bar\Psi(Px)\gamma^\nu\gamma_5\Phi(Px)~, \nonumber\\
        {\cal P}^{-1}\bar\Psi(x)\Sigma^{\mu\nu}\Phi(x){\cal P} &=&  
        + P^\mu_{~\alpha}P^\nu_{~\beta}\bar\Psi(Px)\Sigma^{\alpha\beta}\Phi(Px) ~,
        \label{Dcurrp}
    \eea
where  $\Sigma^{\mu\nu} =\frac{i}{2}[\gamma^\mu,\gamma^\nu]$ are the Lorentz algebra generators. 
Also, here we have used that $\{\gamma_5,D({\cal P})\}=0$, and $\{\gamma^i,D({\cal P})\}=0$, but $[\gamma^0,D({\cal P})]=0$. 

{\bf Under $\cal T$}.
For time reversal we should recall that the transformation is antilinear, and thus, a general bilinear current transforms as ${\cal T}^{-1}\bar\Psi(x){\cal M}\Phi(x){\cal T} = {\cal T}^{-1}\bar\Psi(x){\cal T}{\cal M}^*{\cal T}^{-1}\Phi(x){\cal T} = \bar\Psi(Tx)D({\cal T})^\dagger{\cal M}^* D({\cal T})\Phi(Tx)$, where $\cal M$ 
is any given matrix. 
Applying this rule, using that $D({\cal T})^\dagger(\gamma^\mu)^*D({\cal T}) = (\gamma^\mu)^\dagger = -T^\mu_{~\nu}\gamma^\nu$ and $[\gamma_5,D({\cal T})]=0$, we get
    \bea
        {\cal T}^{-1}\bar\Psi(x)\Phi(x){\cal T} &=&  
        +\bar\Psi(Tx)\Phi(Tx)~,\nonumber\\
        {\cal T}^{-1}\bar\Psi(x)\gamma_5\Phi(x){\cal T} &=&  
        +\bar\Psi(Tx)\gamma_5\Phi(Tx)~,\nonumber\\
        {\cal T}^{-1}\bar\Psi(x)\gamma^\mu\Phi(x){\cal T} &=&  
        -T^\mu_{~\nu}\bar\Psi(Tx)\gamma^\nu\Phi(Tx)~, \nonumber\\
        {\cal T}^{-1}\bar\Psi(x)\gamma^\mu\gamma_5\Phi(x){\cal T} &=&  
        -T^\mu_{~\nu}\bar\Psi(Tx)\gamma^\nu\gamma_5\Phi(Tx)~,\nonumber \\
        {\cal T}^{-1}\bar\Psi(x)\Sigma^{\mu\nu}\Phi(x){\cal T} &=&  
        -T^\mu_{~\alpha}T^\nu_{~\beta}\bar\Psi(Tx)\Sigma^{\alpha\beta}\Phi(Tx)  ~.
        \label{Dcurrt}
    \eea
{\bf ${\cal CPT}$}.
We can combine all transformation rules to show that under $\cal CPT$
    \bea
        ({\cal CPT})^{-1}\bar\Psi(x)\Phi(x){\cal CPT} &=& +\bar\Phi(-x)\Psi(-x)~,\nonumber\\
        ({\cal CPT})^{-1}\bar\Psi(x)\gamma_5\Phi(x){\cal CPT} &=&  -\bar\Phi(-x)\gamma_5\Psi(-x)~,\nonumber\\
        ({\cal CPT})^{-1}\bar\Psi(x)\gamma^\mu\Phi(x){\cal CPT} &=&  
        -\bar\Phi(-x)\gamma^\mu\Psi(-x)~, \nonumber\\
        ({\cal CPT})^{-1}\bar\Psi(x)\gamma^\mu\gamma_5\Phi(x){\cal CPT} &=&  
        -\bar\Phi(-x)\gamma^\mu\gamma_5\Psi(-x)~, \nonumber\\
        ({\cal CPT})^{-1}\bar\Psi(x)\Sigma^{\mu\nu}\Phi(x){\cal CPT} &=&  
        +\bar\Phi(-x)\Sigma^{\mu\nu}\Psi(-x) ~,
        \label{CPTtheo}
    \eea
where we have used that $P^\mu_{~\nu}T^\nu_{~\alpha} = -\delta^\mu_\alpha$.
From these CPT transformations, combined with the transformation rules of the Lorentz tensor fields that should be contracted with the above currents to provide a scalar Lagrangian, it follows that a (local) real Lorentz invariant (fermionic) Lagrangian provides a  CPT invariant action. That is the basis of the so-called CPT Theorem.

\bibliography{ref}

\end{document}